\pdfoutput=1 

\documentclass[aps,prd,reprint,groupedaddress]{revtex4-1}

\usepackage{graphicx}

\usepackage{amsmath}

\begin{document}


\title{Dynamical Pion Collapse and the Coherence of Conventional Neutrino Beams}


\author{B.J.P. Jones}
\email[]{bjpjones@mit.edu}
\affiliation{Massachusetts Institute of Technology}


\date{\today}

\begin{abstract}
In this paper we consider the coherence properties of neutrinos produced by the decays of pions in conventional neutrino beams.   Using a multi-particle density matrix formalism we derive the oscillation probability for neutrinos emitted by a decaying pion in an arbitrary quantum state.  Then, using methods from decoherence theory we calculate the pion state which evolves through interaction with decay-pipe gases in a typical accelerator neutrino experiment.  These two ingredients are used to obtain the distance scales for neutrino beam coherence loss.  We find that for the known neutrino mass splittings, no non-standard oscillation effects are expected on terrestrial baselines.  Heavy sterile neutrinos may experience terrestrial loss of coherence, and we calculate both the distance over which this occurs and the energy resolution required to observe the effect.  By treating the pion-muon-neutrino-environment system quantum mechanically, neutrino beam coherence properties are obtained without assuming arbitrary spatial or temporal scales at the neutrino production vertex.
\end{abstract}

\pacs{}

\maketitle

\section{Introduction}
The standard formula describing neutrino flavor oscillation \cite{Pontecorvo} has been widely experimentally verified \cite{Agashe:2014kda}.  However, it is known that its usual derivation, which proceeds via Taylor expansion of the phases of neutrino mass eigenstates in the plane-wave basis, contains several theoretical inconsistencies and assumptions that are not realized in any experiment.  Attempts to fix these inconsistencies have involved the introduction of internal wave-packets \cite{Kayser:2010pr,Asahara:2004mh,Lychkovskiy:2009uj}, the replacement of internal wave packets with external ones \cite{Akhmedov:2009rb,Akhmedov:2010zz}, field theoretical reformulations \cite{Akhmedov:2010zz,Wu:2010yr,Beuthe:2001rc}, and discussions of the role of the entangled muon in maintaining or suppressing coherence \cite{Boyanovsky:2011xq,Kayser:2010bj,Kayser:2011jn,Wu:2010tr,Wu:2010yr,Cohen:2008qb}.  Many of these approaches turn a poorly formulated quantum mechanical problem into a theoretically robust calculation, but often at the cost of introducing arbitrary spatial or temporal scales.  These scales have yet to be rigorously connected to experiments.  Since several of these approaches predict observable coherence loss effects, especially in experiments searching for hypothetical sterile neutrinos, a robust quantum mechanical understanding of this system without arbitrary scales is needed.

In this paper we present a derivation of the oscillation probability for neutrinos produced in conventional neutrino beams.  In such beams, relativistic pions are injected into a gas-filled decay pipe which is at atmospheric pressure, where they
undergo electromagnetic interactions with gas molecules, eventually decaying to muons and neutrinos.  The muons continue to interact with the environment but are
undetected experimentally, and the neutrinos propagate over some baseline $L$ before being detected via a weak interaction.  The coherence properties of the resulting neutrino beam are influenced by the initial pion state, the interactions of the pion with the decay-pipe environment, the presence of the entangled muon, and the source/detector configuration.

Our approach is to first derive the oscillation probability for neutrinos emitted from pions in an arbitrary initial state, consistently incorporating the constraints imposed by entanglement with an unobserved lepton (Section \ref{sec:ArbPionDecay}).   Then, using a representative pion density matrix, the classical (diagonal) and quantum (off-diagonal) uncertainties on the pion position are shown to lead naturally to two distinct neutrino coherence conditions (Section \ref{sec:CohCond}).  The first mechanism for coherence loss corresponds to a classical smearing of oscillation over the neutrino production point; the second to a distance-dependent suppression of oscillations via wave-packet separation.  If either condition is not satisfied, neutrino oscillations will be suppressed, being replaced by an incoherent mixture of flavors with no nontrivial $L$ or energy dependence.  

With tools from decoherence theory we calculate the pion density matrix which evolves through interactions with the decay-pipe gas, and show how dynamical collapse leads to pion states with a stable and predictable quantum width in position space (Section \ref{sec:PiEvolution}).  This width, in turn, allows the determination of the distance over which oscillations remain observable for a conventional neutrino beam, without requiring the introduction of arbitrary production or detection states (Section \ref{sec:Observability}). 

We then discuss the implications of this result for the oscillations of standard and sterile neutrinos in accelerator neutrino experiments.  We demonstrate that for standard neutrinos, no coherence loss is expected on terrestrial scales.  Effects on terrestrial scales may be present for heavy sterile neutrinos, and we calculate the detector energy resolution required to observe such an effect.  Finally, we discuss the similarities and differences in the coherence properties of other neutrino emission systems (Section \ref{sec:OtherSystems}), and present our conclusions (Section \ref{sec:Conclusion}).

\section{Neutrino Oscillations from Decays of Arbitrarily Prepared Pions \label{sec:ArbPionDecay}}

In this paper we use the density matrix formalism of quantum mechanics \cite{breuer2007theory,schlossauer,NielsenChuang}.   This is equivalent to the more prevalent wave-function formalism, although it accommodates more naturally both quantum and classical superpositions.  This is particularly important in the presence of environmental entanglement, which acts to suppress coherence, effectively converting quantum uncertainties into classical ones within the neutrino subsystem.  Aspects of neutrino oscillations have been analyzed using both wave-function \cite{Kayser:2010bj, Akhmedov:2012uu} and density matrix \cite{Wu:2010tr,Cohen:2008qb} approaches, although no study to date has treated the full pion-lepton-neutrino-environment system.  We find the multi-particle density matrix approach to be a powerful tool for this purpose, allowing treatment of the neutrino beam as an open quantum system and giving new insights into its coherence properties.

We work in a one-dimensional model, beginning with a general pion density matrix in the momentum basis $\rho_\pi$.  This pion, with energy $E_\pi$, may either interact hadronically in the beam-stop or decay with a 99.99\% branching fraction via $\pi \rightarrow \mu \nu_\mu$ in the decay pipe.  In the case where the pion decays, after a sufficiently long time $t$, the two-particle density matrix for the resulting entangled muon / neutrino state can be expressed in the basis of neutrino mass eigenstates $|m_i\rangle$ as:
\begin{equation}
\rho(t)=N^{2}U_{\mu i}U_{\mu j}^{\dagger}\Theta_{ij}(t)|m_{i}\rangle\langle m_{j}|.
\label{eq:numuentangled1}
\end{equation}
In this equation, $U$ is the unitary neutrino mixing matrix and $N$ is a normalization factor.   The momentum degrees of freedom are collected into matrix  $\Theta_{ij}$, shown in (\ref{eq:numuentangled2}).  The functions $p_{\mu}^{i}(p)$ and $p_{\nu}^{i}(p)$ represent the fixed momentum of a muon or neutrino as determined by two-body kinematics in the plane-wave basis.  In each case, $p$ is the momentum of the pion and the decay products have masses $m_\mu$ and $m_i$ respectively.
\begin{widetext}
\begin{eqnarray}
\Theta_{ij}(t)=\int dp_{1}dp_{2}\rho_{\pi}(p_{1},p_{2})e^{i\left(E_{\pi}(p_{1})-E_{\pi}(p_{2})\right)t}\left(|p_{\nu}^{i}(p_{1})\rangle \langle p_{\nu}^{j}(p_{2})|\right)_{\nu}\left(|p_{\mu}^{i}(p_{1})\rangle \langle p_{\mu}^{j}(p_{2})|\right)_{\mu}.
\label{eq:numuentangled2}
\end{eqnarray}
\end{widetext}

To trace out the leptonic degrees of freedom from (\ref{eq:numuentangled2}) we express all of the leptonic states in terms of the basis states $p_{\mu}^{a}(p)$ corresponding to some neutrino mass $m_a$.  These are related by $p_{\mu}^{a}(p_{\pi})=p_{\mu}^{i}(p_{\pi}+2\delta_{\mu}^{ia})$, where,
\begin{subequations}
\begin{eqnarray}
\delta_{\mu}^{ia}=\frac{1}{2}\frac{dp_{\mu}^{i}}{dm_i^2}\left/\frac{dp_{\mu}^{i}}{dp_{\pi}}m_{ia}^{2}\right.
&=-\frac{E_{\pi}}{2M^{2}}m_{ia}^{2},
\label{eq:deltadef1}
\\
M^{2}=&m_{\pi}^{2}-m_{\mu}^{2},
\label{eq:Mdef1}
\end{eqnarray}
\end{subequations}
with $m^2_{ia}=m_i^2-m_a^2$.  The second equality in (\ref{eq:deltadef1}) follows from momentum conservation, and (\ref{eq:Mdef1}) assumes $m_i$ to be much smaller than $m_\pi$ and $m_\mu$.  Expressed in this basis, the muon degrees of freedom can now be traced out of (\ref{eq:numuentangled2}), leading to the reduced density matrix for the neutrino subsystem:
\begin{eqnarray}
\label{eq:ReducedDM} \Theta_{ij}^{\nu}=\int dp\rho_{\pi}(p-\delta_{\mu}^{ij},p+\delta_{\mu}^{ij})e^{i\frac{p}{M^{2}}m_{ij}^{2}t} \times & \nonumber
\\
|p_{\nu}^{i}(p-\delta_{\mu}^{ij})\rangle \langle p_{\nu}^{j}(p+&\delta_{\mu}^{ij})|.
\end{eqnarray}

This object can be used to obtain the results of any measurement which can be performed on the neutrino alone, and incorporates the effects of both the initial pion state and the unmeasured entangled muon.  An example of such a measurement is a neutrino oscillation experiment.  A positive operator valued measure (POVM) is applied which selects a particular neutrino flavor state at baseline L, giving (\ref{eq:OscProb}), the probability for neutrino detection in flavor $\alpha$:
\begin{equation}
P(\nu_{\alpha})=U_{\alpha j}U_{\alpha i}^{\dagger}U_{\mu i}U_{\mu j}^{\dagger} 
\int dq_{1}dq_{2}\langle q_{1}|\Theta_{ij}^{\nu}|q_{2}\rangle e^{i(q_{1}-q_{2})L} \label{eq:OscProb}.
\end{equation}
where summation over $i$ and $j$ is implicit.  Substituting (\ref{eq:ReducedDM}) into (\ref{eq:OscProb}) leads to the final expression for the generalized oscillation probability,

\begin{gather}
P(\nu_{\alpha},L)=U_{\alpha j}U_{\alpha i}^{\dagger}U_{\mu i}U_{\mu j}^{\dagger}\int dp\rho_{\pi}(p-\delta_{\mu}^{ij},p+\delta_{\mu}^{ij})e^{-i\phi_{ij}}\nonumber \\
\phi_{ij}=\frac{m_{ij}^{2}p}{M^{2}}\left(\frac{E}{p}L-t\right)\label{eq:GeneralNeutrinoOsc}.
\end{gather}
We note that this expression does not depend on the muon final state, or its subsequent interactions.  A general proof that this must be the case is presented in Appendix \ref{ap:Lepton}.

\section{Pion State Coherence Conditions \label{sec:CohCond}}

Equation (\ref{eq:GeneralNeutrinoOsc}) allows us to calculate the neutrino oscillation probability from any pion, prepared in a general state of coherent or incoherent superposition.  In order to explore the coherence properties of such a system, we consider as an example a pion density matrix in the position basis, with Gaussian diagonal width $\sigma_{diag}$, off-diagonal width $\sigma_{od}$, and central momentum $p_{0}$:
\begin{gather}
\rho=\int dx_1 dx_2 \: \mathrm{exp}\left[-\frac{\left(x_1 - x_2\right)^{2}}{2\sigma_{od}^{2}}-\frac{\left(x_1 + x_2 \right)^{2}}{2\sigma_{diag}^{2}}\right] \; \times \nonumber \\ 
e^{{-ip_{0} \left(x_1-x_2 \right)}}|x_1\rangle\langle x_2|\label{eq:ChoiceofDM}.
\end{gather}
The diagonal width corresponds to the classical uncertainty on the pion position, whereas the off-diagonal width corresponds to a quantum mechanical, coherent uncertainty.  Substituting (\ref{eq:ChoiceofDM}) into (\ref{eq:GeneralNeutrinoOsc}), after a few lines of algebra we can acquire the oscillation probability in the relativistic $(p_0 \gg m_\pi)$ and non-relativistic $(p_0 \ll m_\pi)$ pion limits, as (\ref{eq:GeneralizedProbR}), (\ref{eq:GeneralizedProbNR}):

\begin{widetext}
\begin{subequations}
\begin{eqnarray}
P_{R}(\nu_{\alpha},L)&=&NU_{\alpha j}U_{\alpha i}^{\dagger}U_{\mu i}U_{\mu j}^{\dagger}\left[e^{i\frac{m_{ij}^{2}}{2p_{0}^{2}}\frac{m_{\pi}^{2}}{m^{2}-m_{\mu}^{2}}L}\right]\left[e^{-\frac{(m_{ij}^{2})^{2}(p_{0}^{2}+m_{\pi}^{2})}{8(m_{\pi}^{2}-m_{\mu}^{2})^{2}}\sigma_{daig}^{2}}\right]\left[e{}^{-\left(\frac{m_{ij}^{2}}{m_{\pi}^{2}-m_{\mu}^{2}}\frac{m_{\pi}^{2}}{2p_{0}^{2}}\right)^{2}\left(\frac{L^{2}}{2\sigma_{od}^{2}}\right)}\right]\
\label{eq:GeneralizedProbR}
\\
P_{NR}(\nu_{\alpha},L)&=&NU_{\alpha j}U_{\alpha i}^{\dagger}U_{\mu i}U_{\mu j}^{\dagger}\left[e^{i\frac{m_{ij}^{2}m_{\pi}}{m_{\pi}^{2}-m_{\mu}^{2}}\left(1+\frac{p_{0}}{m_{\pi}}\right)^{-1}L}\right]\left[e^{-\frac{(m_{ij}^{2})^{2}(p_{0}^{2}+m_{\pi}^{2})}{8(m_{\pi}^{2}-m_{\mu}^{2}){}^{2}}\sigma_{daig}^{2}}\right]\left[e^{-\left(\frac{m_{ij}^{2}}{m_{\pi}^{2}-m_{\mu}^{2}}\right)^{2}\left(1+\frac{p_{0}}{m_{\pi}}\right)^{-2}\left(\frac{L^{2}}{2\sigma_{od}^{2}}\right)}\right]
\label{eq:GeneralizedProbNR}
\end{eqnarray}
\end{subequations}
\end{widetext}

Describing the bracketed factors of (\ref{eq:GeneralizedProbR}) and (\ref{eq:GeneralizedProbNR}) from left to right: 
The left factor contains the standard oscillation phase, reported here as a function of the central
pion momentum $p_{0}$. The central factor describes incoherent
smearing caused by the production of neutrinos
over a non-negligible spatial region. This is significant
if the neutrino production region $\sigma_{diag}$
is larger than the neutrino oscillation length.  As well as a position-space interpretation, this
effect also has an interpretation in the momentum basis, as the requirement that momentum wave-packets
for different mass states should overlap in order for oscillations to occur.
The rightmost factor accounts
for position-dependent wave packet separation, which is a function
of the off-diagonal width $\sigma_{od}$ and becomes more severe with
increasing $L$.  Equation (\ref{eq:GeneralizedProbR}), with which we will be primarily concerned, reduces to
the standard formula for neutrino oscillations in the limits:
\begin{subequations}
\begin{eqnarray}
\sigma_{diag}&\ll&\left(\frac{E_{\pi}^{0}}{2\sqrt{2}\left(m_{\pi}^{2}-m_{\mu}^{2}\right)}m_{ij}^{2}\right)^{-1}
\label{eq:cohcond1}
\\
\frac{\sigma_{od}}{L}&\gg&\frac{1}{2\sqrt{2}p_{0}^{2}}\frac{m_{\pi}^{2}}{\left(m_{\pi}^{2}-m_{\mu}^{2}\right)}m_{ij}^{2} \label{eq:cohcond2}
\end{eqnarray}
\end{subequations}
which can be considered as ``coherence conditions'' for neutrino oscillations to be observable.  

Condition (\ref{eq:cohcond1}), representing classical smearing of the oscillation, has been discussed at length as a source of incoherence, for example \cite{Akhmedov:2012uu,Hernandez:2011rs}.  However, since the majority of experiments necessarily account for this classical averaging in their Monte Carlo simulations, it will not introduce unexpected effects for properly simulated neutrino experiments, so we will not discuss it further.  

Condition (\ref{eq:cohcond2}), on the other hand, is a constraint on the quantum mechanical width of the wave packet that dictates the distances over which wave packets for different mass eigenstates become separated spatially. This term can modify the standard neutrino transformation probability at large distances,  and does not have a classical interpretation.  Constraints similar to  (\ref{eq:cohcond2}) have been derived for the special case of a pure coherent state with $\sigma^2_{od}=\sigma^2_{diag}=2 \sigma^2_x$ \cite{Beuthe:2002ej,Akhmedov:2012uu}.  However, the pion state which evolves in a typical neutrino beam is neither a pure state nor a minimal-uncertainty one, and the width $\sigma_{od}$ has not been rigorously determined.

\section{Evolution of the Pion State in a Conventional Beam-line \label{sec:PiEvolution}}

We now turn to the determination of the pion state at neutrino production. In several prior studies of this system, a Gaussian state was assumed with width
related to a physical scale in the problem, and the position and momentum space widths related by the minimal uncertainty relation $\sigma_x \sigma_p = \hbar / 2$.  Interactions with decay-pipe gas are typically neglected.  Following the method developed in \cite{Tegmark:1993yn}, we will show that the decay-pipe gas interactions are crucial to the coherence properties of the neutrino beam system.  This is because the bombardment of a particle by environmental scatters leads to a highly squeezed state with a predictable quantum width. 
This ``dynamical collapse'' is experienced by the pion sufficiently early in its lifetime that all neutrinos can be assumed to be emitted from a pion
with a stable equilibrium width.  

\begin{figure}
\includegraphics[width=0.99\columnwidth]{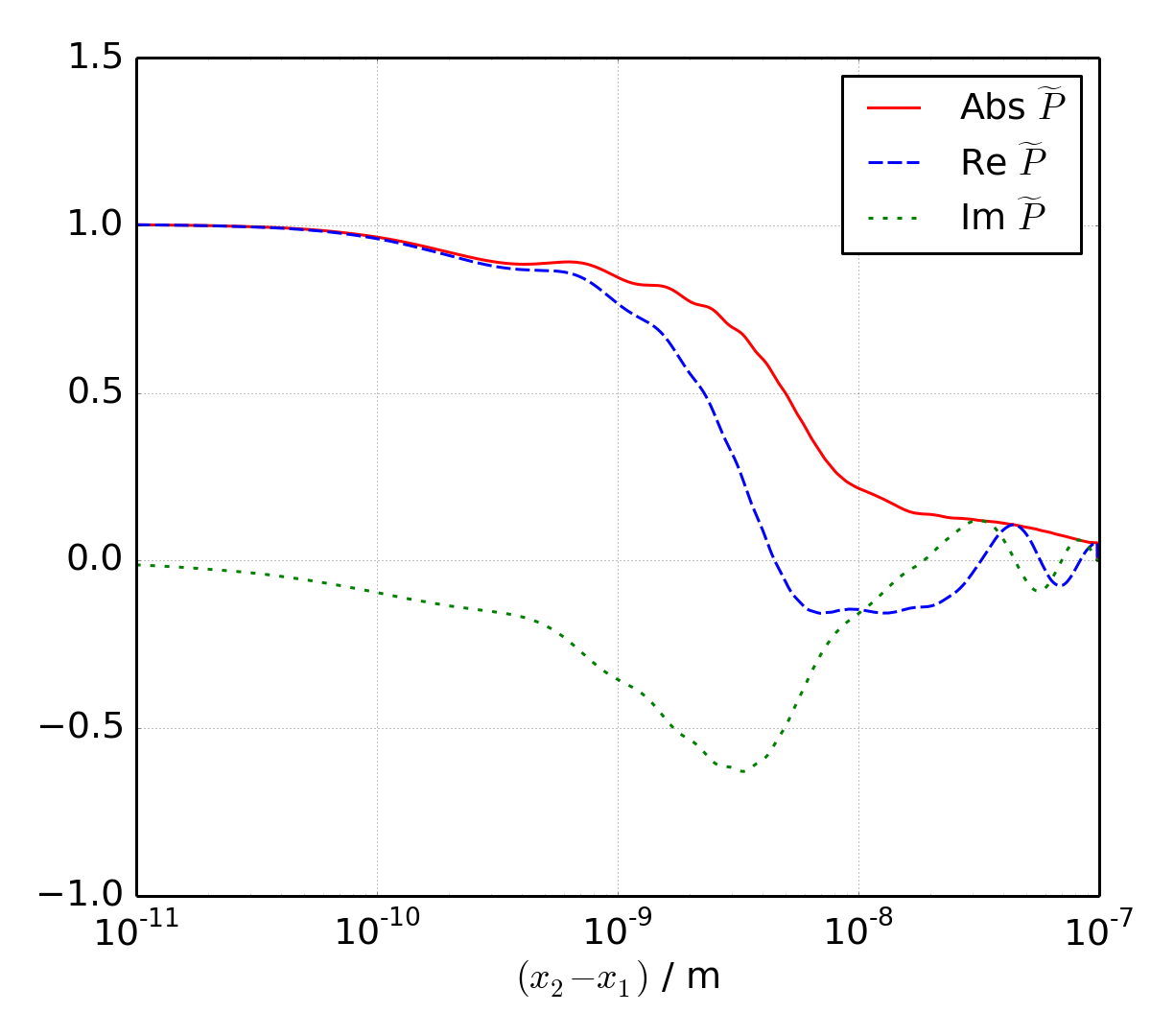}%
\\
\includegraphics[width=0.97\columnwidth]{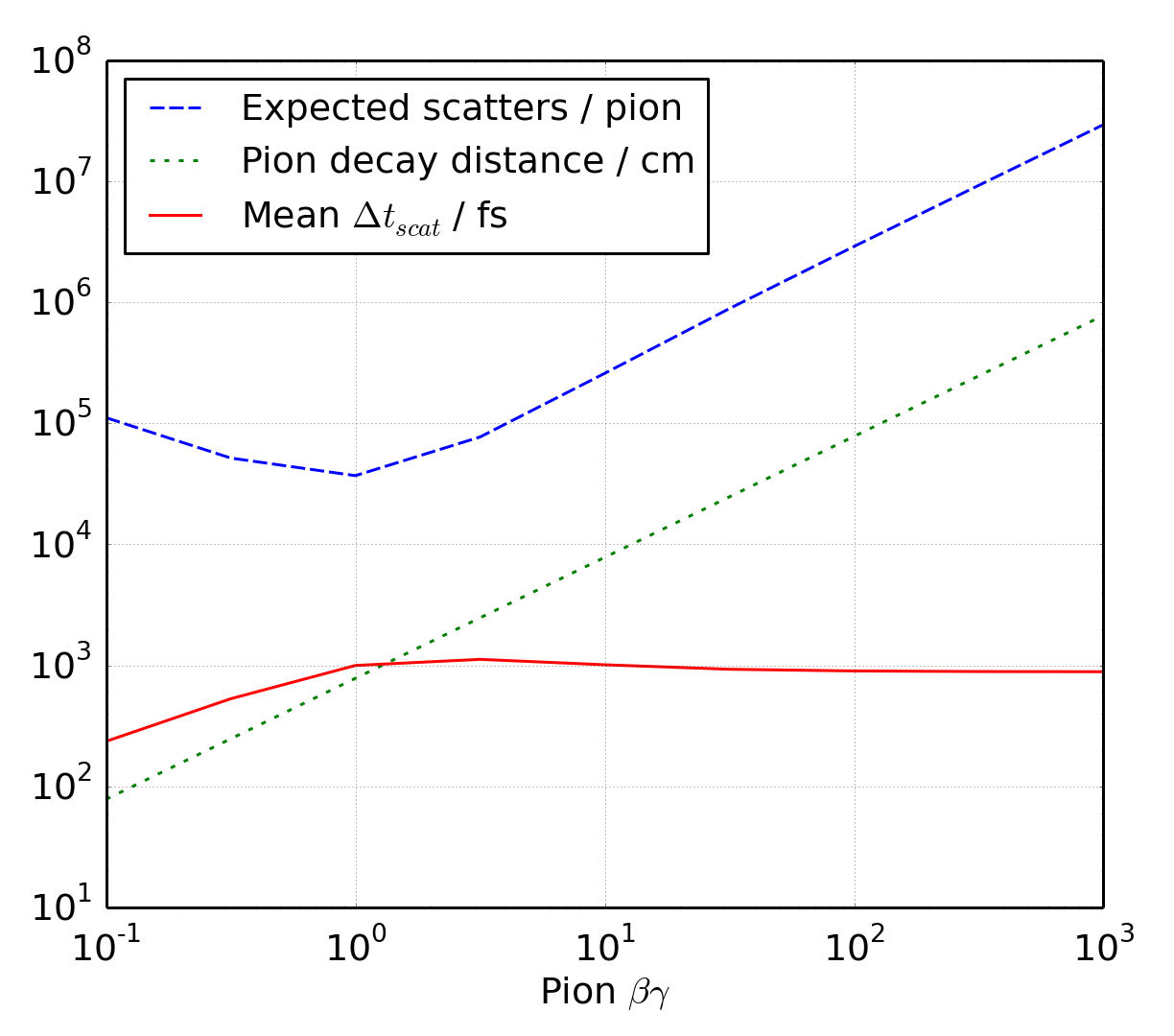}%
\caption{\label{fig:DecFunc} Predictions of the PAI model. Top : the decoherence function for the pion in interaction with decay-pipe gas for $\beta\gamma=10$.  Bottom: pion scattering rate and other related quantities for pions of different momenta}
\end{figure}

Since we cannot track every air molecule in the decay-pipe,
the evolving pion is treated as an open quantum system. Air molecules undergoing local
 interactions with the pion encode information
about its position into the environment, causing an effective collapse
of its wave function in the position basis. As demonstrated in Ref. \cite{Tegmark:1993yn}, the collapsing resolution of the environment is determined by the scale 
of momentum transfers.  The effect of a single
scatter with momentum transfer probability distribution $P(q)$ on the pion density matrix is  

\begin{equation}
\rho_{\pi}(x_1,x_2)\rightarrow \rho_{\pi}(x_1,x_2)\widetilde{P_{q}}(x_2-x_1),
\end{equation}
with $\widetilde{P_{q}}(x_2-x_1)$ being the Fourier transform of $P(q)$.  This relationship was derived for non-relativistic scatterers, but remains valid for the relativistic pion.
Competing with the effective collapse caused by scattering is wave packet dispersion, which acts to broaden the wave packet in position space between scatters. The competition between these two processes leads to a stable coherent width, which depends on the pion energy in two key ways. 1) the rate of dispersion for more energetic pions is suppressed by a Lorentz factor $\gamma$, and 2) the rate and momentum distribution of environmental scatters depends upon the pion energy.  

To obtain the probability distribution of momentum transfers $P(q)$ we use the PAI model \cite{Allison:1980vw}.  This model uses classical electromagnetism 
to determine the energy losses in a continuous medium parameterized
by a complex index of refraction. The energy loss incorporates
both ionization and Cherenkov losses, which are the dominant sources of
momentum transfer for relativistic particles traversing matter at these energies. The imaginary part
of the refractive index is related to the photo-absorption cross section
of the material in the VUV range, and the real part is obtained via the Kramers Konig relations. The continuous energy loss is then re-interpreted semi-classically in terms of discrete
photon exchanges with electrons to give a distribution of momentum transfers. This
distribution has been used to predict the fluctuations in $dE/dx$ of
ionizing particles in drift chambers and good agreement
with experimental data is observed \cite{Allison:1980vw}.

Since photoionization data for nitrogen are not available at the level of detail required, we use the photoionization spectrum for argon gas as input, making the assumption that the details of
atomic shell structure will not cause large differences in the shape of the momentum transfer distribution.  Predictions of our PAI model implementation were checked against those given in the original paper \cite{Allison:1980vw}, and total $dE/dx$ as reported in the PDG \cite{Agashe:2014kda}, with good agreement observed in both cases.  More information on our implementation of the PAI model and cross-checks is given in Appendix \ref{ap:PAI}.

The required outputs of the PAI model are a) the complex decoherence function $\widetilde{P_{q}}(x_2-x_1)$ and b) the scattering rate, both functions of pion energy and shown in Figure \ref{fig:DecFunc}.  These are then used in a quantum Monte Carlo simulation of the dynamical collapse of the pion state.

We perform this Monte Carlo on a gridded space of dimension $2048\times 2048$.
The resolution of this space in the momentum basis $r_{p}$ is related to its resolution in the position basis by $r_{p}=2\pi/r_{x}D$, specified for each initial state such that it is equally sized in the position and momentum grids at $t = 0$. The center of the grid corresponds initially to $x_{G}=0$ and $p_{G}=p_{0}$, with these values updating as the grid moves to track the center of the wave-packet. 

We construct an initial Gaussian pure state of width $\sigma_{initial}$ and central momentum $p_{0}$ in this space. The following procedure is then applied:
\begin{enumerate}
\item Sample the time-to-next-interaction, $t_{evol}$ from a distribution $P(t_{evol})=(\Delta t_{scat})^{-1} e^{-t_{evol}/\Delta t_{scat}(p_{0})}$, where $\Delta t_{scat}(p_0)$ is the mean scattering time (Figure \ref{fig:DecFunc}, bottom);
\item In the momentum basis, unitarily evolve the state for a time $t_{evol}$;
\item Fourier transform the density matrix into the position basis;
\item Apply the decoherence function $\tilde{P}(x_2-x_1)$, an example of which is shown in Figure \ref{fig:DecFunc}, top.
\item Fourier transform back to the momentum basis;
\item Continue until the density matrix overflows the grid boundaries in one of
the two spaces.
\end{enumerate}

 \begin{figure}
\includegraphics[width=0.99\columnwidth]{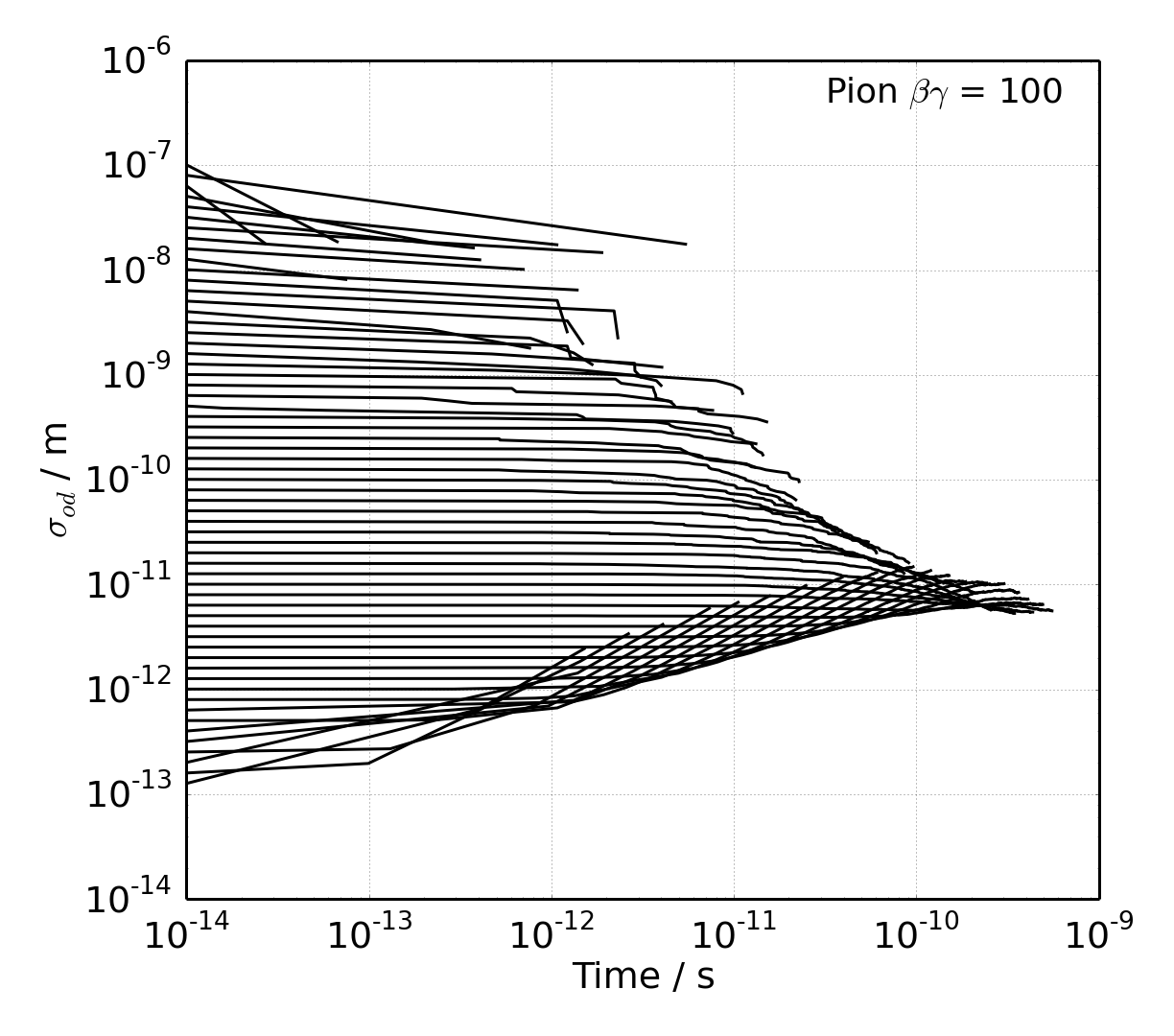}%
\caption{\label{fig:EvolBG100}The time-dependent coherent position-space width of pions with $\beta\gamma=100$.  Gaussian pure states of various widths are used as initial states in a quantum monte-carlo.  Each simulation is stopped when the wave packet no longer fits on the simulation grid in the position or momentum basis.  Evolution towards a stable equilibrium width is observed.}
\end{figure}

 \begin{figure}
\includegraphics[width=0.99\columnwidth]{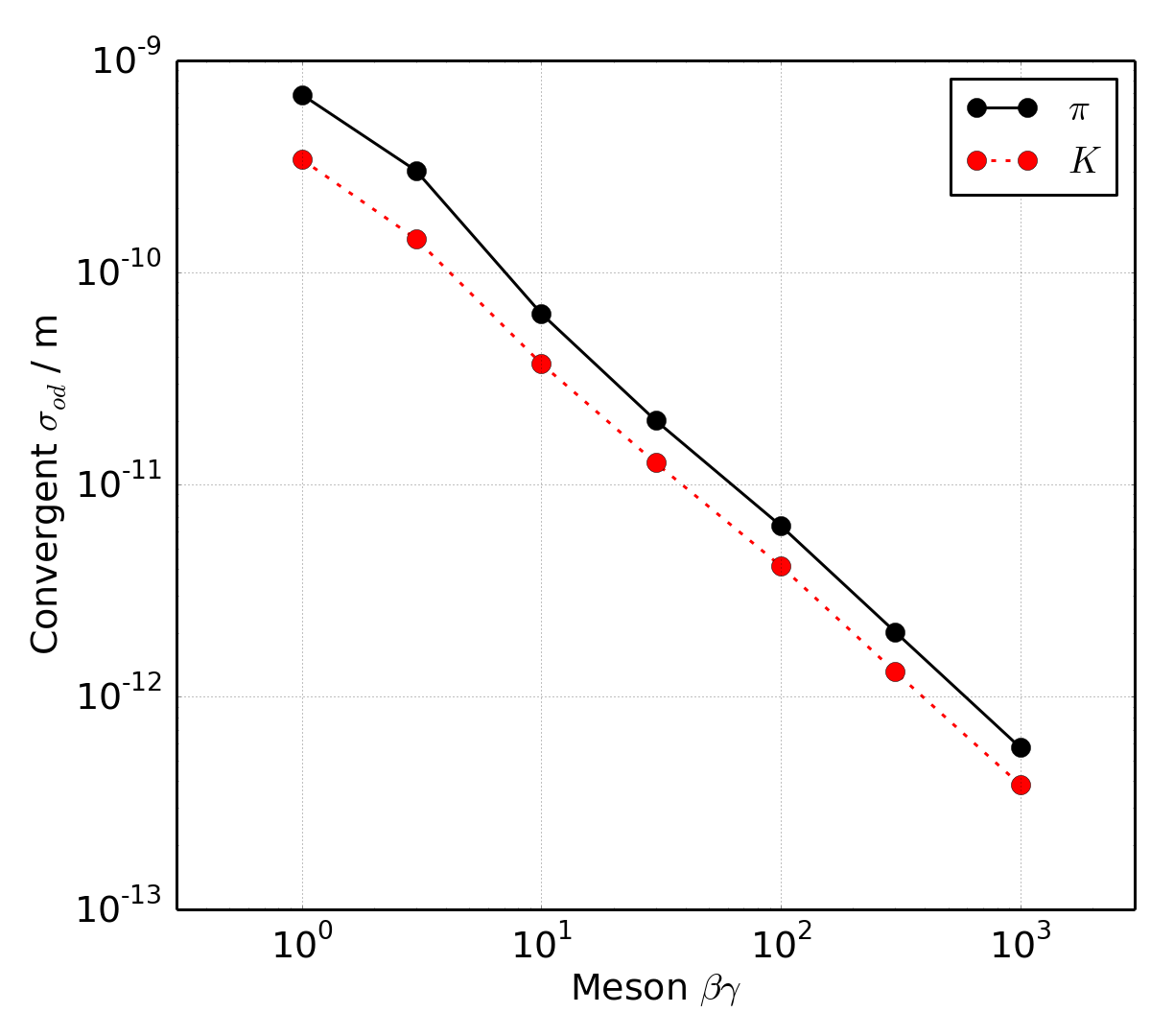}%
\caption{\label{fig:ConvergentWidths} Convergent widths for pions and kaons of different initial energies}
\end{figure}

 \begin{figure}
\includegraphics[width=0.99\columnwidth]{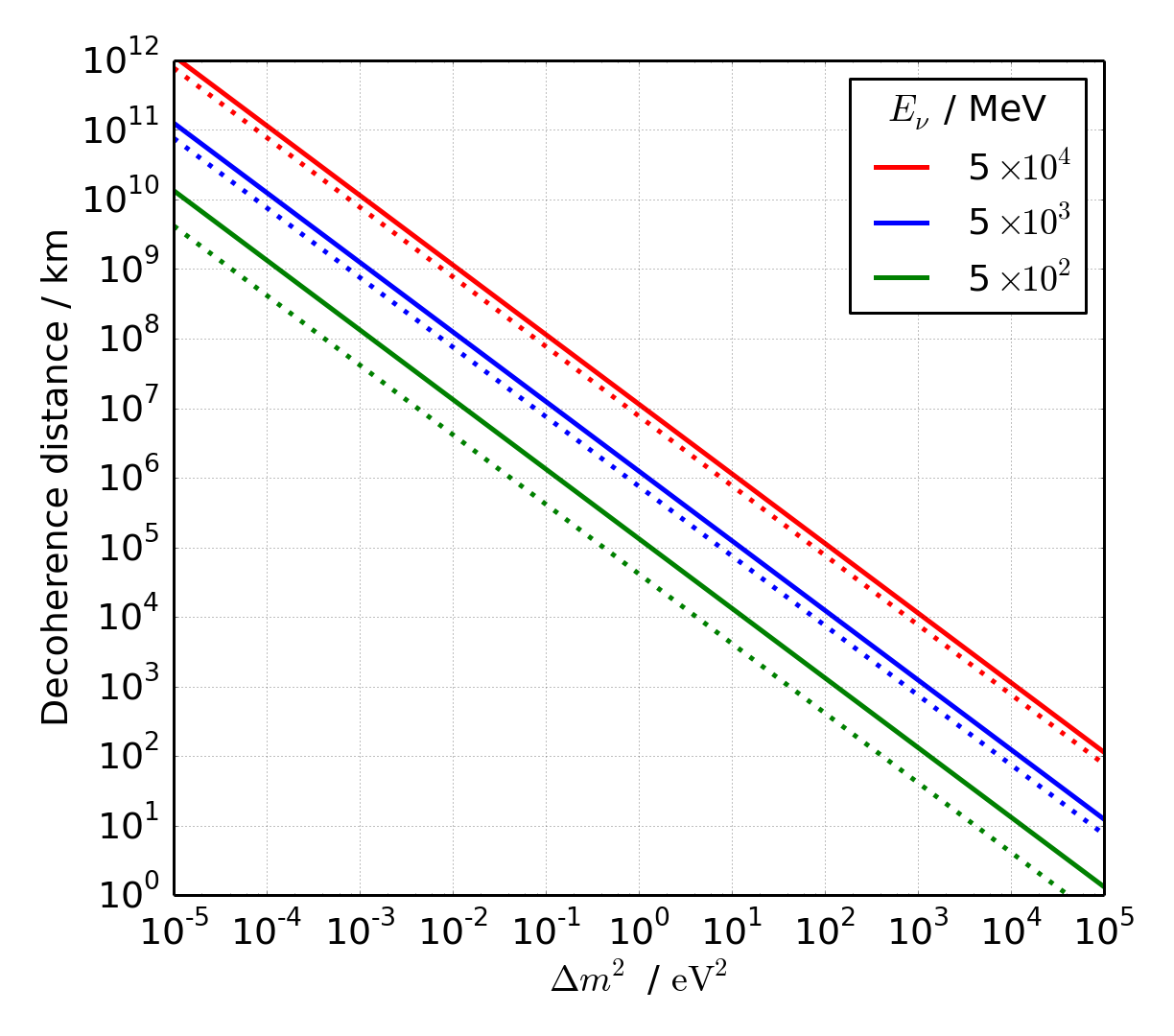}%
\caption{\label{fig:DecoherenceLength} Coherence distance for neutrinos of different energies produced in conventional neutrino beams. Solid : Two body $\pi^\pm$ decay.  Dotted : Two body $K^\pm$ decay.}
\end{figure}

After each interaction we record the diagonal and
off-diagonal position- and momentum-space widths by calculating the
standard deviation of the density matrix at its peak in the diagonal
and off-diagonal directions. The calculation is halted when any width
reaches 1 grid spacing, or when the amplitude of the wave-packet at
any edge of the grid is more than 1\% of its peak.  Further cross-checks of the simulation can be found in
Appendix \ref{ap:CrossChecks}.  An example of the evolution of the coherent pion width
 for various initial states at $\beta\gamma=100$ is shown in Figure \ref{fig:EvolBG100}, where the convergence to an equilibrium width is clearly observed.  A similar calculation can be performed for kaons, whose two-body decays give a subdominant flux contribution to conventional neutrino beams, by substitution $m_\pi\rightarrow m_K$.  The asymptotic widths for pions and kaons of different energies are shown in Figure \ref{fig:ConvergentWidths}.

\section{Coherence of Pion Beams and Observability of Coherence Loss Effects \label{sec:Observability}}

The asymptotic coherent pion and kaon widths shown in Figure \ref{fig:ConvergentWidths} can be used with coherence condition (\ref{eq:cohcond2}) to determine the distance over which the pion- and kaon-induced fluxes in conventional neutrino beams will become incoherent due to wave-packet separation.  This distance is shown as a function of neutrino mass splitting in an effective two neutrino system with $m^2_{ij}=\Delta m^2$, for several energy points in Figure \ref{fig:DecoherenceLength}.  

\begin{figure}
\includegraphics[width=0.99\columnwidth]{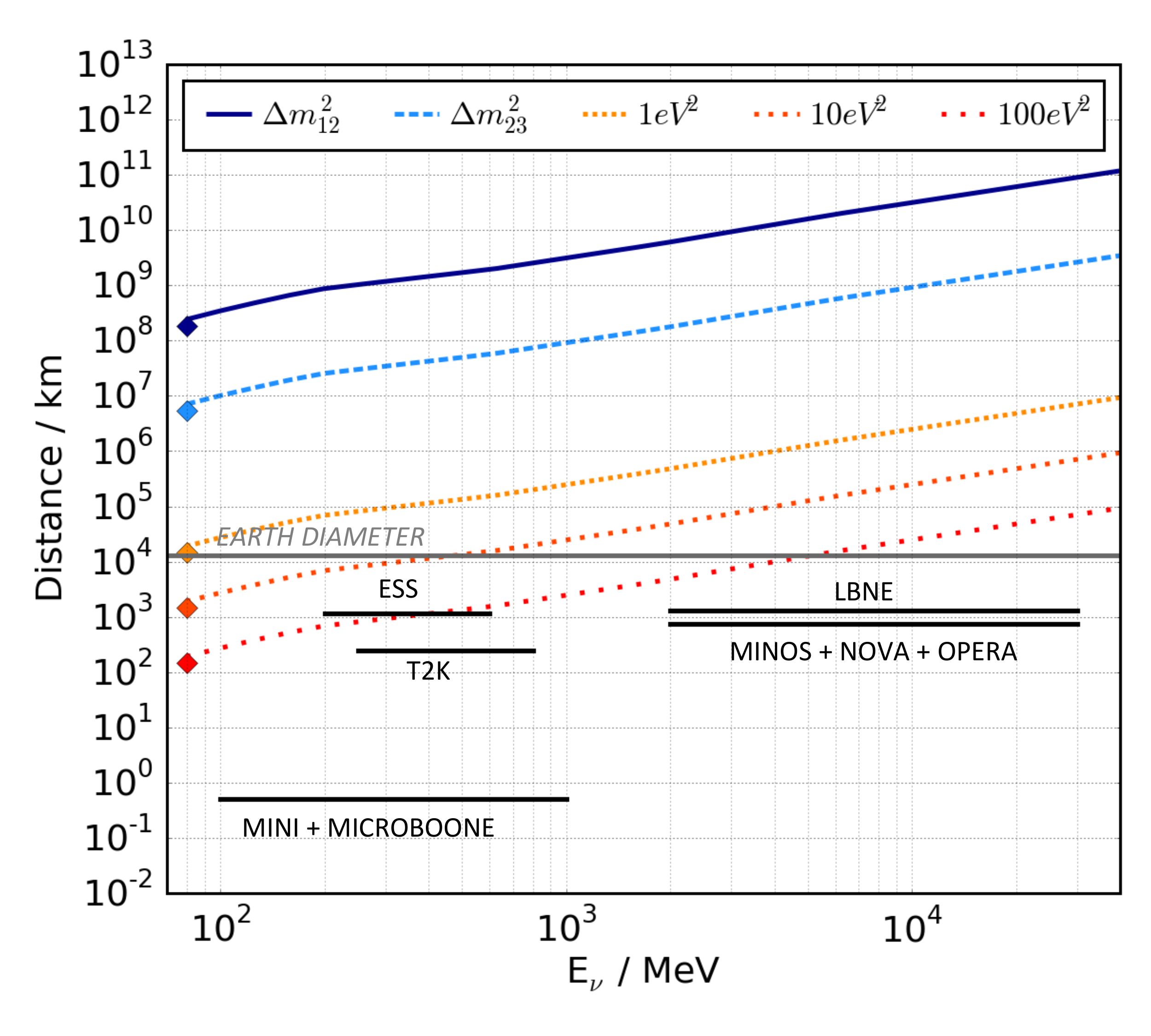}%
\caption{\label{fig:DecoherenceVsEnergy} The coherence distances for pion-induced neutrinos at several values of $\Delta m^2$ compared with the configurations of existing and proposed accelerator neutrino experiments.  The lines use the relativistic expression (\ref{eq:GeneralizedProbR}), whereas the diamonds show the prediction of the non-relativistic expression (\ref{eq:GeneralizedProbNR}) at the lowest energies.}
\end{figure}

Existing and near-future accelerator neutrino experiments have baselines of up to 1300 km and beam energies from $10^2$ to $10^5$ MeV.  Figure \ref{fig:DecoherenceVsEnergy} shows the energy range and baseline for several such experiments, as well as the predicted coherence distance for several mass splittings, calculated using the relativistic formula (\ref{eq:GeneralizedProbR}).  The lowest energy point on this plot corresponds to a pion energy of 210 MeV, where non-relativistic corrections may be expected.  To illustrate the scale of such corrections we show the prediction of non-relativistic expression  (\ref{eq:GeneralizedProbNR}) at this point.  We observe that the discrepancy with the relativistic prediction is small at these energies. 

Over most of the energy range the coherence distance scales with $E$.  This can be understood as emerging from the factor of $p_0^2$ in equation (\ref{eq:cohcond2}) combined with the characteristic $p_0^{-1}$ scaling of the convergent width, caused by Lorentz suppression of the pion dispersion rate.  At low energies there are corrections both from the finite pion mass and from the energy dependence of the mean scattering time shown in Figure \ref{fig:DecFunc}.

It is clear from Figure \ref{fig:DecoherenceVsEnergy} that no loss of coherence expected for the standard neutrinos produced in conventional beams on terrestrial distance scales.  However, effects may be present for heavy sterile neutrinos.  The experimental observability of these effects is directly related to the energy resolution of the far detector.  At large $L$ and $\Delta m^2$, the oscillation phase $\Delta m^2 L / E$ varies rapidly as a function of energy.  If this variation is more rapid than the energy resolution of the detector, an incoherent signal and a fast-oscillating one are indistinguishable.

\begin{figure}
\includegraphics[width=0.99\columnwidth]{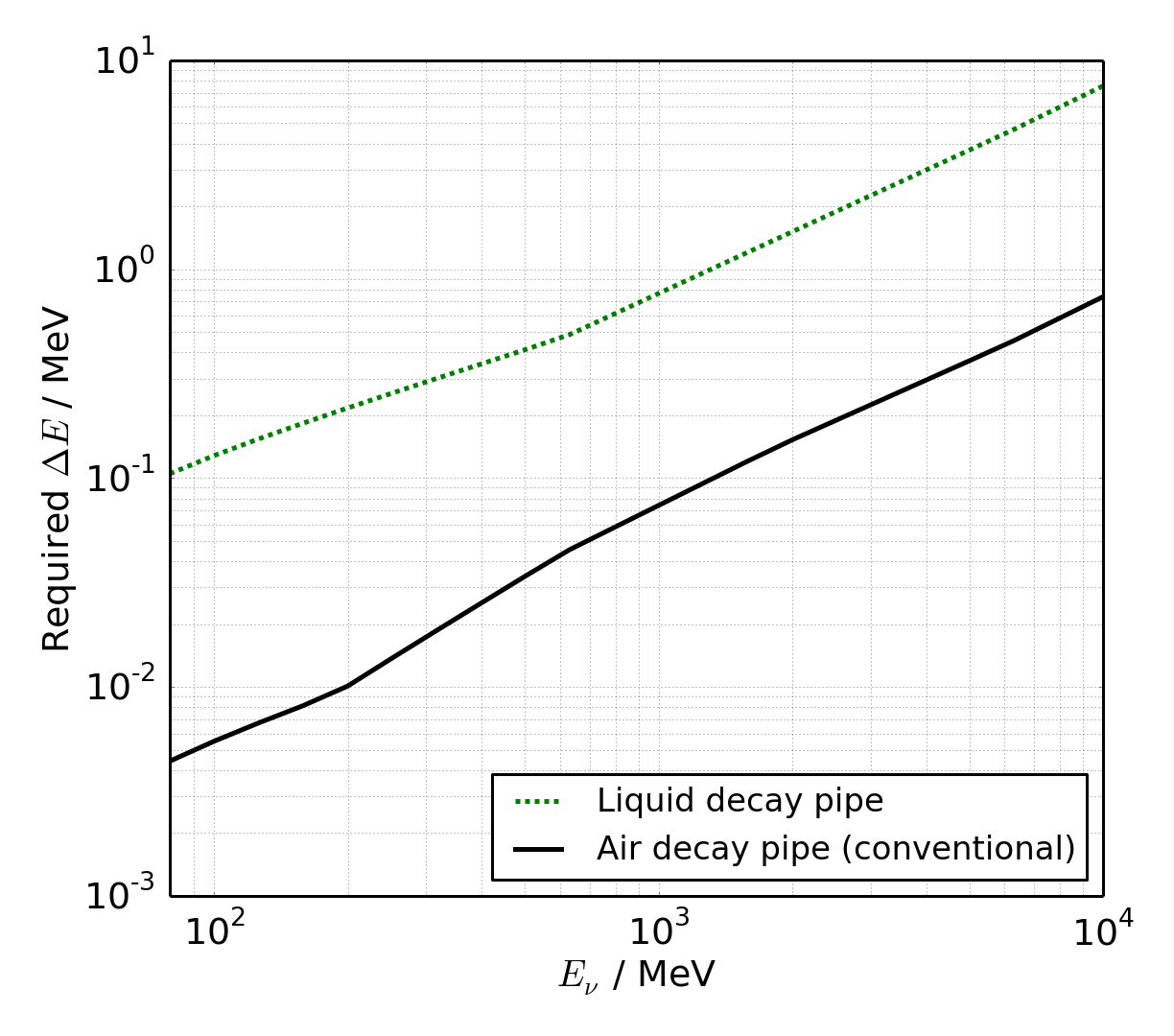}%
\caption{\label{fig:EnergyResolution} The energy resolution required to observe coherence loss without fast-oscillation in some part of the $\Delta m^2 L$ parameter space as a function of neutrino energy.}
\end{figure}

Taylor expanding the oscillation phase $\Delta m^2 L / (E+\Delta E)$ in the limit of small $\Delta E / E$ gives the required energy resolution $\Delta E$ for observability of oscillations:

\begin{equation}
\Delta E < \frac{2\pi E^2}{\Delta m^2 L} \label{eq:EnergyRes}.
\end{equation}
Comparison of (\ref{eq:EnergyRes}) with (\ref{eq:cohcond2}) gives the range of $\Delta m^2 L$ over which loss of coherence both occurs and is experimentally observable.  For such a space to exist at all, the energy resolution must satisfy:
\begin{equation}
\Delta E < \frac{\pi}{\sqrt{2}\sigma_{od}(E)}\frac{m^2_\pi}{m^2_\pi-m^2_\mu} \label{eq:ResCondition}
\end{equation}
This condition is shown in Figure \ref{fig:EnergyResolution}.  In the range of terrestrial loss of coherence, a resolution  $ < 0.1$ MeV is required. 

It may be possible to loosen this requirement with exotic experimental configurations.  For example, if the pion traverses a decay volume with a higher density, the time between scatters will be reduced and the convergent wave-packet width will be narrowed, leading to enhanced position-space loss of coherence at shorter baselines.  Consider a \textit{Gedankenexperiment} where instead of air, a relativistic pion decays whilst traversing a high-density liquid.  Producing such a beam in practice would introduce a myriad of experimental problems due to rapid energy loss of pions in the medium, but it provides an illustration of the scale of effects which can in principle be probed by changing the decay pipe density.  Using the PAI model for liquid argon, which is 1200 times more dense than air, we re-calculate convergent wave packet widths and the required energy resolutions for the observability of coherence loss, which are shown in Figure \ref{fig:EnergyResolution}.  We see that although the energy resolution condition is relaxed, it remains very precise by existing neutrino detection standards.

\section{Coherence of Other Neutrino Oscillation Systems \label{sec:OtherSystems}}

This paper has focussed on the coherence properties of conventional meson-decay-in-flight neutrino beams.  However, this is only one sub-class of neutrino oscillation experiment, with others using different neutrino sources with different coherence properties.  In this section we give a qualitative discussion of the similarities and differences between some of these systems.

\textit{Three-body decay-in-flight} of muons in conventional neutrino beams requires a small modification of the calculation we have presented.  The coherent width of the muon at the time of decay can be calculated using our method, and since its mass is similar to that of the pion, we expect to find similar coherent widths.  The decay then proceeds $\mu\rightarrow e\nu\bar{\nu}$.  To calculate the oscillation probability for the neutrino, for example, both the electron and the antineutrino must be traced out of the multi-particle density matrix.  The momenta of the unobserved subsystem can be re-parameterized in terms of a total momentum and an invariant mass $m_{\mu\bar{\nu}}$, which implies that this system can be treated in our formalism as a neutrino recoiling against a variable-mass muon.  Since each configuration of the antineutrino-muon subsystem is distinct, the neutrinos recoiling from different invariant masses will not interfere coherently.  Thus the final expression for the oscillation probability will take the form of an integral of Eq. (\ref{eq:GeneralNeutrinoOsc}) over a probability distribution of $m_\mu = m_{\mu\bar{\nu}} $ values, with this distribution calculated from 3-body kinematics.   The coherence properties are likely to be quite similar to those of the pion-decay system.  

\textit{Muon storage rings}, on the other hand, produce neutrinos through the decays of muons which are circulated in an evacuated beam-pipe to maintain a long storage time.  In this case, ionizing interactions with residual gases may no longer be the dominant localizing influence, and our method of calculating the convergent wave-packet width using the PAI model is unlikely to be directly applicable.

\textit{Atmospheric neutrinos} are primarily produced in the decays of charged pions and kaons in the upper atmosphere, with pions dominant at lower and kaons dominant at higher energies.  The quantum mechanical system is essentially identical to that presented in this note, although the atmospheric density at the altitude where air-showers develop is much less than that in accelerator neutrino beam decay pipes.  Our calculation could be applied by simply adjusting the density of the decay-pipe gas to the atmospheric density, giving wider wave-packets and longer coherence lengths.  Because the expected effect is coherent broadening, wave packet separation effects are unlikely to be experimentally observable for atmospheric neutrinos in Earth-based experiments.

\textit{Pion decay-at-rest beams} present a similar quantum mechanical system to pion-decay-in-flight, with the steps leading to Eq. (\ref{eq:GeneralNeutrinoOsc}) remaining valid. A different calculation is required for the initial pion width, since the momentum transfers involved in localizing a stopped particle are characteristically different to those produced by ionizing interactions in the relativistic case.  Stopped $\pi^-$ become trapped in atomic orbitals and capture on nuclei, so do not produce effective decay-at-rest beams.  Stopped $\pi^+$ do not become trapped in atomic orbitals and become stopped somewhere in the material lattice until they decay.  Electromagnetic and phonon interactions with the surrounding material are the primary localizing influence.  A derivation of the coherence properties of decay-at-rest beams would require a microscopic model of the momentum transfers involved in these interactions with the pion to determine its coherent width at decay.

\textit{Reactor and solar neutrinos} are produced by nuclear $\beta$ decays of atoms in a hot, dense environment.  The final state contains a daughter nucleus and an electron.  In the density matrix formalism, the degrees of freedom of final state which are not carried by the neutrino should be traced out to obtain the neutrino reduced density matrix.  Accounting for all of the internal and external degrees of freedom of the daughter nucleus may require fairly involved nuclear physics.  The localization of the initial state is also nontrivial, having contributions from the atomic interactions in the hot medium, photon exchange between the nucleus and its electron cloud, and Fermi motion within the nucleus itself.  Although the latter certainly involves the largest momentum transfers, it is not conceptually clear what role is played by localization of the nucleon within the nuclear medium versus localization of the decaying nucleus within the bulk in determining the coherence properties of the emitted neutrino.  This is a system whose coherence properties warrant further study, and where our calculation, though giving some insight, cannot be trivially applied. \\

\section{Conclusion \label{sec:Conclusion}}

Using a multi-particle density matrix formalism we have derived an expression for the oscillation probability of neutrinos produced by the two-body decays of pions in an arbitrary initial state.  Assuming an example initial state with specified diagonal and off-diagonal widths in the position basis, we derived two coherence conditions for the observability of neutrino oscillations  which set an upper limit on the classical width and a lower limit on the quantum mechanical width respectively.  Modeling the dynamical collapse of a pion in a beam pipe using tools from decoherence theory and the PAI model to obtain realistic momentum-transfer distributions, we calculated the coherence distances for neutrinos produced in conventional neutrino beams.  To our knowledge, this is the first calculation to consistently treat the full pion-muon-neutrino-environment system.  We find that no coherence loss should be expected for standard neutrinos on terrestrial scales in existing or proposed facilities.  Sterile neutrinos with large masses ($>$ 10 $\mathrm{eV} ^2$) at low energies and long baselines may lose coherence through wave-packet separation on terrestrial scales, although a far detector with better energy resolution than is presently available is likely to be required in order to observe this effect.

\begin{acknowledgments}
The author thanks Janet Conrad, Andr\'{e} de Gouv\^{e}a, Boris Kayser and Joachim Kopp for discussions and guidance at all stages of this work, and particularly JC and AdG for the time they dedicated to helping prepare this manuscript.  Thanks also to Jonathan Asaadi, Eric Church, Gabriel Collin, Yoni Kahn, Ben Safdi, Josh Spitz, and Jesse Thaler for a thorough reading of this paper and for valuable comments. The author is supported by National Science Foundation Grant PHY-1205175.
\end{acknowledgments}

\clearpage

\begin{appendix}
\section{The Effects of Muon-Environment Interactions \label{ap:Lepton}}
Here we give a proof that the environmental interactions or detection state of the entangled lepton cannot affect the oscillation phenomenology of a neutrino beam.  Whilst this conclusion is not difficult to reach intuitively via causality arguments, it is one which is either unclear or explicitly violated by several treatments of the neutrino beam system which can be found in the literature, so a formal derivation is useful.

Consider the muon, neutrino, environment system after the pion decay
has taken place. This muon-neutrino subsystem is in an entangled state
represented by reduced density matrix $\rho_{\mu\nu}$. We assume
arbitrary entanglement at $t_{0}$, but make a Schmidt decomposition
into the $\epsilon$ (environment) and entangled $\mu\nu$ (muon and neutrino)
subspaces (choosing a basis for each subsystem such that the entanglement is diagonal):

\begin{equation}
\rho(t_{0})=\rho_{\mu\nu\epsilon}(t_{0})=\sum_{i}\lambda_{i}\rho_{\mu\nu}^{i}\otimes\rho_{\epsilon}^{i} \label{eq:Ap1InitState}
\end{equation}
This density matrix will evolve into another fully entangled state as $ \rho(t_{0})\rightarrow\rho(t)=\rho_{\mu\nu\epsilon}(t) $.  At any time we can obtain the reduced neutrino density matrix from the full
density matrix by tracing out the other degrees of freedom:

\begin{equation}
\rho_{\nu}(t)=\mathrm{Tr}_{\epsilon\mu}\left[\rho_{\mu\nu\epsilon}(t)\right]
\end{equation}

Any measurement we make on the neutrino alone can be represented by a 
POVM on the neutrino Hilbert space $\left\{ O_{\nu}^{i=1...N}\right\} $, giving probability 
 $P(o_j)=\mathrm{Tr}_{\nu}\left[\rho_{\nu}O_{\nu}^{j}\right]$ of measuring outcome  $o_j$, where
$\sum_{i=1}^{N}O_{\nu}^{i}=1_{\nu}$.

To time-evolve the system we apply the relevant muon-neutrino-environment
Hamiltonian. We assume no interactions between the neutrino and the environment,
although our conclusions remain valid even in the presence
of neutrino-environment interactions, so long as the
degrees of freedom coupling to the neutrino are different
from those coupling to the entangled lepton. The muon and neutrino move apart from the origin and are assumed
not to interact with one another after production. The muon will in
general have an interaction with the environment. Finally all three parts have free evolution Hamiltonians.
Therefore, the total Hamiltonian can be written:

\begin{eqnarray}
H=&H_{\nu}&+H_{\mu}+H_{\epsilon}+H_{\mu\epsilon}^{Int},  \nonumber \\
=&H_{\nu}&+H_{\mu\epsilon}.
\end{eqnarray}

In the second line we have separated the Hamiltonian into a neutrino-only
part and a muon-environment part. These act on different Hilbert spaces
so will always commute, $[H_{\nu},H_{\mu\epsilon}]=0$. The time evolution
operator for the entire system is:

\begin{equation}
\mathcal{U}(t-t_{0})=T\left[exp\left\{ i\int dt\left(H_{\nu}+H_{\mu\epsilon}\right)\right\} \right], \nonumber 
\end{equation}

\begin{equation}
=T\left[exp\left\{ i\int dtH_{\nu}\right\} \right]T\left[exp\left\{ i\int dtH_{\mu\epsilon}\right\} \right], \nonumber 
\end{equation}

\begin{equation}
=\mathcal{U}_{\nu}(t-t_{0})\mathcal{U}_{\mu\epsilon}(t-t_{0}),
\end{equation}
where the second equality is valid because of the commutation properties
already mentioned. Using this operator we evolve the initial density matrix:

\begin{equation}
\rho(t)=\mathcal{U}^{\dagger}\rho(t_{0})\mathcal{U},
\end{equation}
dropping the arguments of the time evolution operators
for simplicity of notation. We substitute the initial state (\ref{eq:Ap1InitState}), 
and use the separability of the time
evolution operator to find:

\begin{equation}
=\sum_{i}\lambda_{i}\mathcal{U}_{\mu\epsilon}^{\dagger}\left[\mathcal{U}_{\nu}^{\dagger}\rho_{\mu\nu}^{i}(t_{0})\mathcal{U}_{\nu}\right]\otimes\rho_{\epsilon}^{i}(t_{0})\mathcal{U}_{\mu\epsilon}
\end{equation}
which leads to (\ref{eq:Ap1GeneralMeasurement}), the general expression for measurement probabilities for any neutrino observable at any time.

\begin{widetext}

\begin{equation}
P(o_{j},t)=\mathrm{Tr}_{\epsilon\mu}\left[\mathrm{Tr}_{\nu}[\rho O_{\nu}^{j}]\right]=\sum_{i}\lambda_{i}\mathrm{Tr}_{\epsilon\mu}\left[\mathcal{U}_{\mu\epsilon}^{\dagger}\mathrm{Tr}_{\nu}\left[\mathcal{U}_{\nu}^{\dagger}\rho_{\mu\nu}^{i}(t_{0})\mathcal{U}_{\nu}O_{\nu}^{j}\right]\otimes\rho_{\epsilon}^{i}(t_{0})\mathcal{U}_{\mu\epsilon}\right] \label{eq:Ap1GeneralMeasurement}
\end{equation}
\end{widetext}

We know that time evolution in quantum mechanics is unitary. This does not necessarily imply unitary evolution within a subsystem, however.  The overall unitarity requirement gives:

\begin{subequations}
\begin{eqnarray}
\mathcal{U}^{\dagger}\mathcal{U}&=&1_{\nu\mu\epsilon}\\
&=&1_{\nu}\otimes1_{\mu\epsilon} \\
&=&\mathcal{U}_{\nu}\mathcal{U}_{\mu\epsilon}\mathcal{U}_{\mu\epsilon}^{\dagger}\mathcal{U}_{\nu}^{\dagger}
\end{eqnarray}
\end{subequations}
We know that time evolution for the neutrino subsystem is unitary, since it is identical
to the free neutrino case. Therefore the states $U_{\nu}|\nu_{i}>$
make just as good orthonormal basis states for taking a trace as $|\nu_{i}>$,
and $\mathrm{Tr}_{\nu}[\mathcal{U}_{\nu}^{\dagger}A_{\mu\nu\epsilon}\mathcal{U}_{\nu}]=\mathrm{Tr}_{\nu}[A_{\mu\nu\epsilon}]$,
where $A_{\mu\nu\epsilon}$ is any general operator on the 3 system
Hilbert space. This tells us that the muon-environment time evolution
must also be unitary in its own Hilbert space $\mathcal{U}_{\mu\epsilon}\mathcal{U}_{\mu\epsilon}^{\dagger}=1_{\mu\epsilon}$, leading to the conclusion

\begin{equation}
\mathrm{Tr}_{\mu\epsilon}\left[\mathcal{U}_{\mu\epsilon}^{\dagger}A_{\mu\nu\epsilon}\mathcal{U}_{\mu\epsilon}\right]=\mathrm{Tr}_{\mu\epsilon}\left[A_{\mu\nu\epsilon}\right]
\end{equation}
We can use this to simplify the above expression (\ref{eq:Ap1GeneralMeasurement}) for neutrino measurement
probabilities:

\begin{equation}
P(o_{j})=\sum_{i}\lambda_{i}\mathrm{Tr}_{\epsilon\mu}\left[\mathrm{Tr}_{\nu}\left[\mathcal{U}_{\nu}^{\dagger}\rho_{\mu\nu}^{i}(t_{0})\mathcal{U}_{\nu}O_{\nu}\right]\otimes\rho_{\epsilon}^{i}(t_{0})\right].
\end{equation}
The operator $\mathcal{U}_{\mu\epsilon}$ no longer features in this expression. We conclude that any measurement made on the neutrino alone cannot be influenced by the subsequent evolution or environmental interactions of the entangled muon. 

Correlated measurements on both subsystems would involve a measurement
operator $O_{\mu\nu}$. In this case, the muon-environment interactions
could, of course, affect the oscillation probabilities of neutrinos
detected in coincidence with a particular subset of muons. This would
correspond to a selection effect, with measurements which are made
over all neutrinos, without reference to their associated muons, still
returning values which are independent of the muon-environment interactions.
Thus, by virtue of quantum mechanical unitarity, there is no spooky-action-at-a-distance
or faster-than-light communication, and the interactions of muons
with their environment cannot affect the neutrino oscillation probability
in any experiment.

\section{Construction of the PAI Model \label{ap:PAI}}
There are two main sources of environmental interaction for pions
in a neutrino beam-line. The first is interaction with the magnetic
fields of the focussing horn which steers the pions into a forward
beam. In the lab frame, this field is purely magnetic, and so only
transfers transverse momentum, having no decohering effect
in the longitudinal direction important for neutrino beam coherence.
Furthermore, the pion is only in the region of strong horn fields
for a short time, spending most of its decay time traveling forward
in the low field region.

The second source of environmental interaction are ionization losses
and Cherenkov emission due to photon exchange with the beam-pipe gas. The
energy loss of a relativistic particle passing through a gaseous environment
is understood as being the result of many low energy scatters with
nuclei and electrons. It is well known that the majority
of the energy loss occurs through scattering with electrons rather
than nuclei, the latter giving a contribution on the order of $O(10^{-4})$
to $dE/dx$. In what follows we will neglect the effects of scatters
off nuclei completely, although a more complete treatment might include
this small correction.

\begin{figure}
\begin{centering}
\includegraphics[width=1.0\columnwidth]{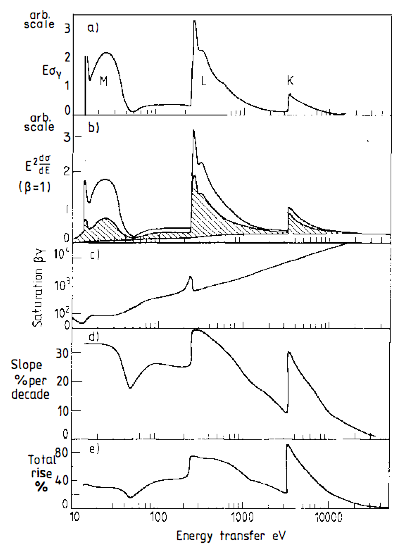}
\par\end{centering}

\caption{Elements of the PAI model. The two panels of interest to us are the
top panel, showing the photo-absorption cross section of argon gas,
and the second panel, showing the calculated $dE/dx$ from this model
for $\beta=1$. This figure is reproduced from \cite{Allison:1980vw}\label{fig:Elements-of-the}}
\end{figure}

The distribution of momentum transfers for relativistic particles in matter
has been studied in several contexts, including
for the purpose of understanding ionization fluctuations in drift
chambers. Therefore much of the existing data and modeling has focused on
common drift chamber gases such as argon and methane.

One example of such a model is the photo-absorpative ionization (PAI)
model \cite{Allison:1980vw}. In this model, classical electromagnetism
is used to determine the energy losses in a continuous medium parameterized
by a complex index of refraction $n$. This energy loss incorporates
contributions from both ionization and Cherenkov emission, which
although both included, are not easily separable. The refractive index is 
determined from the photo-absorption spectrum
of the material in the VUV range, which has been measured for argon
and is shown in figure \ref{fig:Elements-of-the}, reproduced from \cite{Allison:1980vw}. The continuous
energy loss is re-expressed semi-classically in terms of discrete
photon exchanges with electrons to give a distribution of momentum transfers.

The main result we need from the PAI model is Eq. (\ref{eq:PAIEnergyLoss}),
giving the energy transfer cross section in terms of the photo-absorption
cross section of the medium. In this formula, $\beta$ is the ionizing
particle velocity, E is the energy transfer by a single photon exchange,
and $\sigma_{\gamma}(E)$ is the photoionization cross section of
the material in the VUV range.  $\epsilon_{1}$ and $\epsilon_{2}$ are the real and imaginary parts
of the material index of refraction, which can be expressed in terms
of the photo absorption cross section via the definition of the absorption
length, and then the Kramers Konig relation, as in Eqs (\ref{eq:PAIDef1}), (\ref{eq:PAIDef2}). The function $\Theta$ is defined in terms of the 
particle velocity and the index of refraction as in Eq (\ref{eq:PAIDef3}).

\begin{figure}[b]
\begin{centering}
\includegraphics[width=1.0\columnwidth]{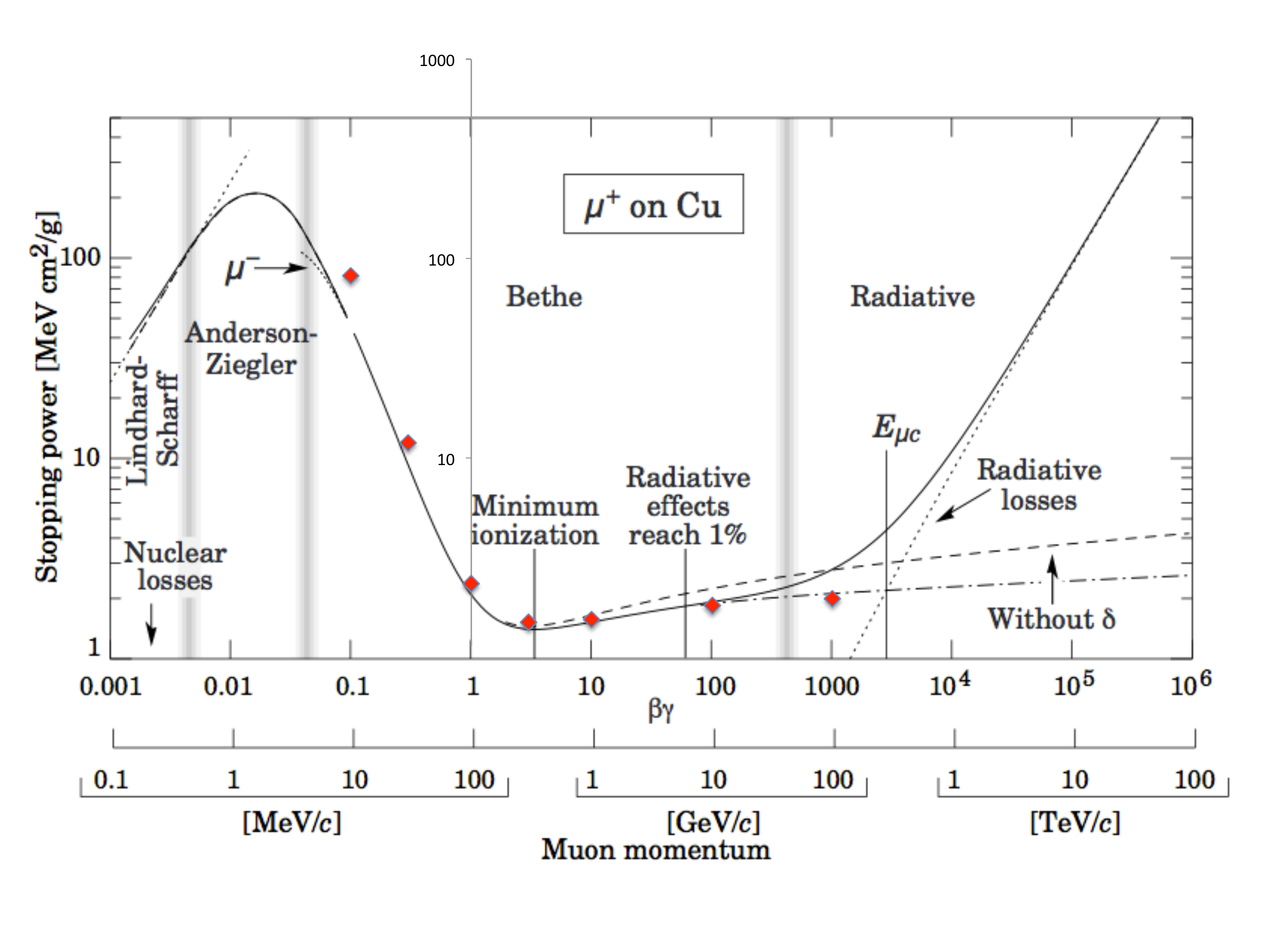}
\par\end{centering}
\caption{dEdx calculated using PAI model (red points) overlaid on the prediction 
from \cite{Agashe:2014kda}.
 \label{fig:dEdxCrossCheck}}
\end{figure}

\begin{widetext}

\begin{equation}
\frac{d\sigma}{dE}=\frac{\alpha}{\beta^{2}\pi}\frac{\sigma_{\gamma}(E)}{EZ}ln\left[(1-\beta^{2}\epsilon_{1})^{2}+\beta^{4}\epsilon_{2}^{2}\right]^{-1/2}+\frac{\alpha}{\beta^{2}\pi}\frac{1}{N\hbar c}\left(\beta^{2}-\frac{\epsilon_{1}}{|\epsilon|^{2}}\right)\Theta \nonumber
\end{equation}

\begin{equation}
+\frac{\alpha}{\beta^{2}\pi}\frac{\sigma_{\gamma}(E)}{EZ}ln\left[\frac{2mc^{2}\beta^{2}}{E}\right]+\frac{\alpha}{\beta^{2}\pi}\frac{1}{E^{2}}\int_{0}^{E}\frac{\sigma_{\gamma}(E')}{Z}dE' \label{eq:PAIEnergyLoss}
\end{equation}

\end{widetext}

\begin{subequations}
\begin{equation}
\epsilon_{2}=\frac{Nc}{\omega Z}\sigma_{\gamma}(E) \label{eq:PAIDef1}
\end{equation}
\begin{equation}
\epsilon_{1}=\frac{2}{\pi}\frac{Nc}{Z}P\int_{0}^{\infty}\frac{\sigma_{\gamma}(x)dx}{x^{2}-\omega^{2}} \label{eq:PAIDef2}
\end{equation}
\begin{equation}
\Theta=arg(1-\epsilon_{1}\beta^{2}+i\epsilon_{2}\beta^{2}) \label{eq:PAIDef3}
\end{equation}
\end{subequations}

\begin{figure}
\begin{centering}
\includegraphics[width=1.0\columnwidth]{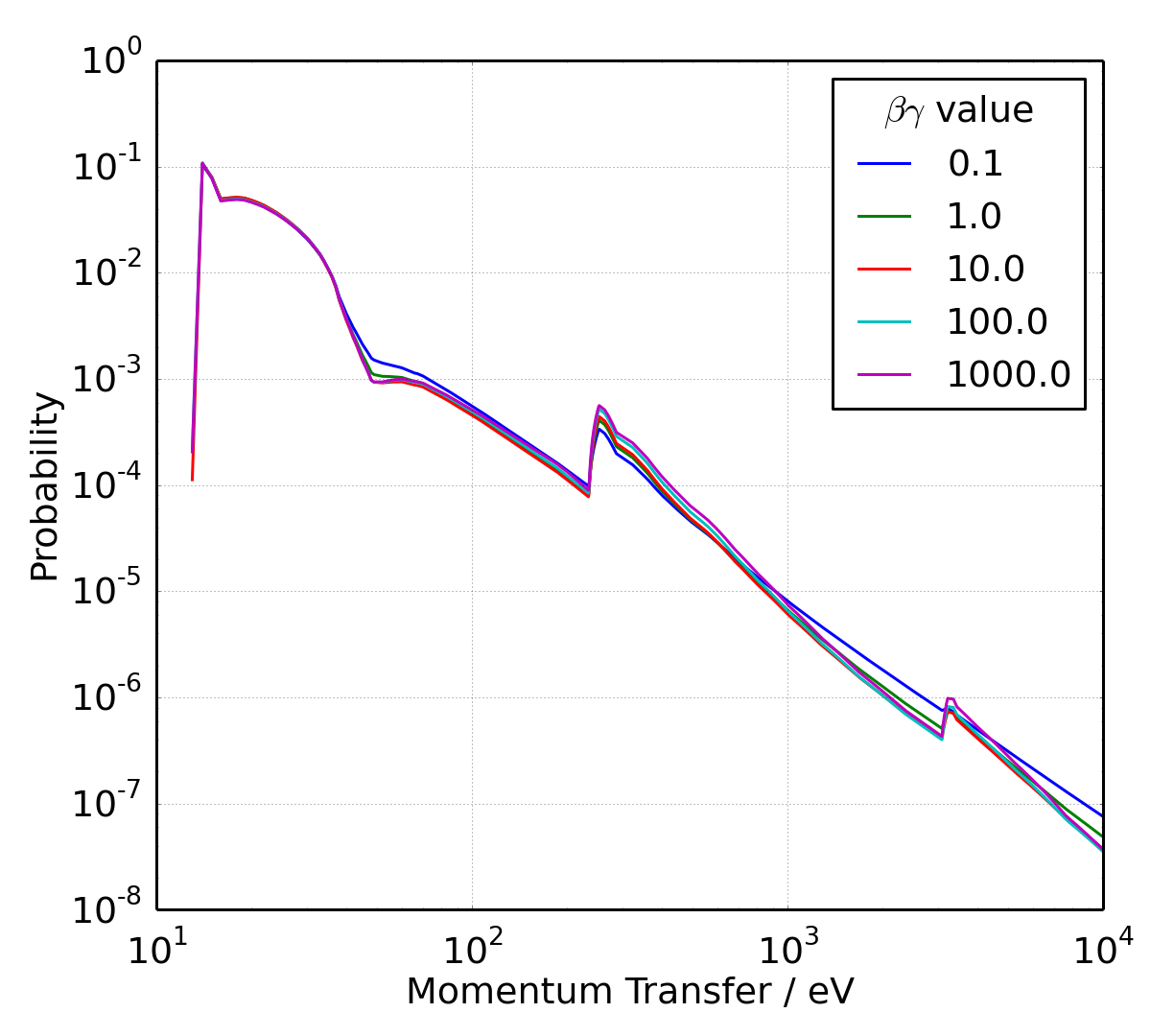}\\
\includegraphics[width=0.98\columnwidth]{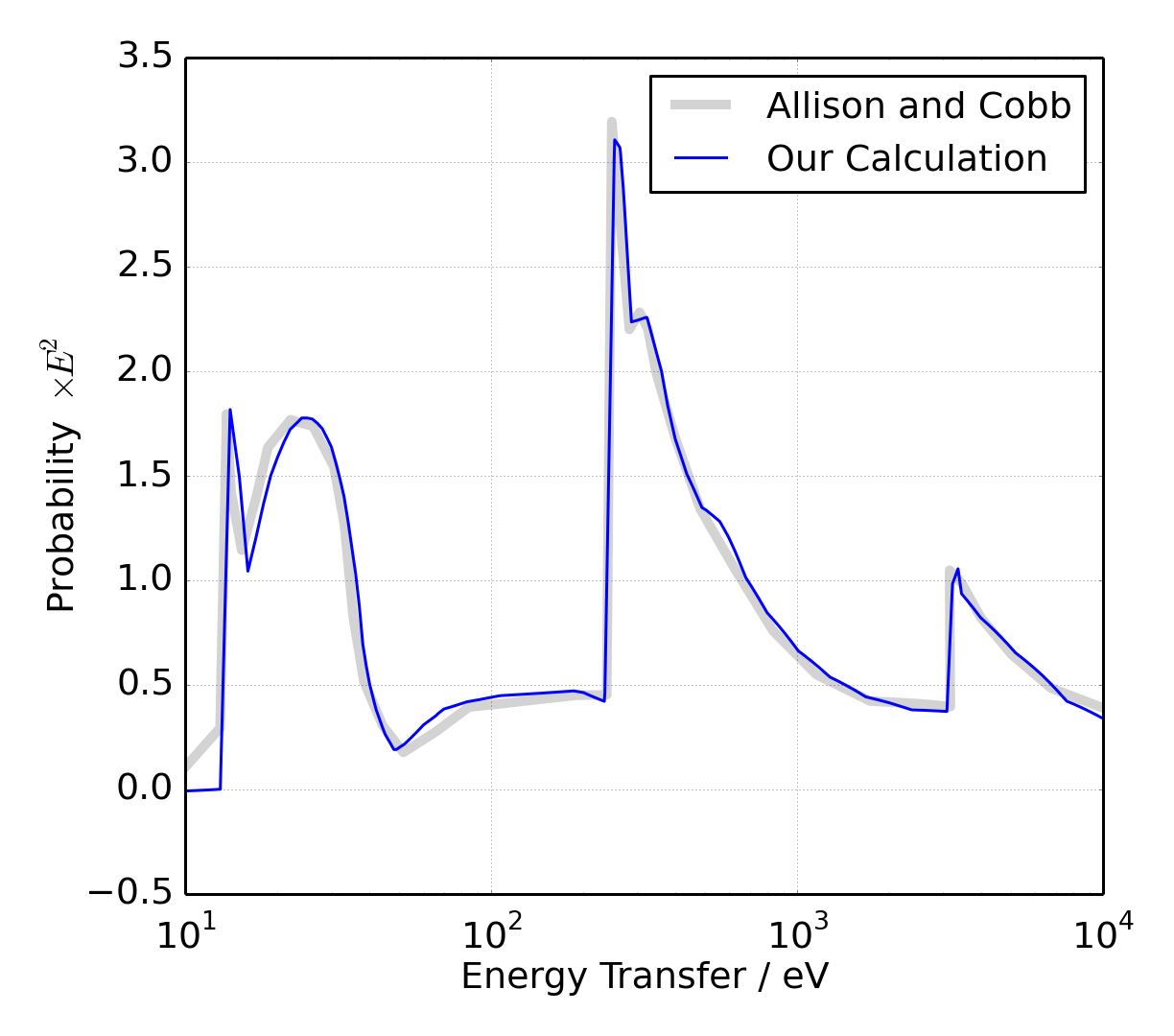}
\par\end{centering}

\caption{
Top: The momentum transfer distribution predicted by the PAI model for various values of $\beta\gamma$. 
Bottom: The energy transfer distribution calculated in our PAI model implementation,
compared with that given in \cite{Allison:1980vw} \label{fig:EnergyTransferCrosscheck}}

\end{figure}

The momentum
transfer probability distribution, and the number of scatters per
centimeter are given by

\begin{equation}
P(q)=\frac{1}{\sigma}\frac{d\sigma}{dp}\qquad\frac{dN_{scat}}{dx}=\frac{1}{<E>}\frac{dE}{dx}
\end{equation}
and the normalized probability distribution for each discrete momentum
transfer at several $\beta\gamma$ values is shown in Figure \ref{fig:EnergyTransferCrosscheck}, top.

We cross-check
our implementation of this model in two ways. First, Figure \ref{fig:EnergyTransferCrosscheck}, bottom
shows our calculated momentum transfer distribution for $\beta=1$
as compared to that given in \cite{Allison:1980vw}. There is good
agreement everywhere except at the lowest momentum transfers, where
it appears the Allison and Cobb calculation displays a small tail which
is not present in our model. We suspect this could be the result of
the choice of regularization in the Fourier transform. Since these
are scatters with very small momentum transfer and so little decohering
resolution, we do not expect this feature to be important for this
calculation. Second, we calculate the total $dE/dx$ predicted by
this model as a function of pion energy, and compare to the PDG. This
is shown in Figure \ref{fig:dEdxCrossCheck}, and again, good agreement
is observed.

\section{Tests of Numerical Density Matrix Evolution \label{ap:CrossChecks}}

Here we present some crosschecks of the density matrix evolution
calculation which test that the convergent
width depends only on the physics of the problem, and not the details
of our simulation. 

\begin{widetext}

\begin{figure}[p]
\begin{centering}
\includegraphics[width=0.8\columnwidth]{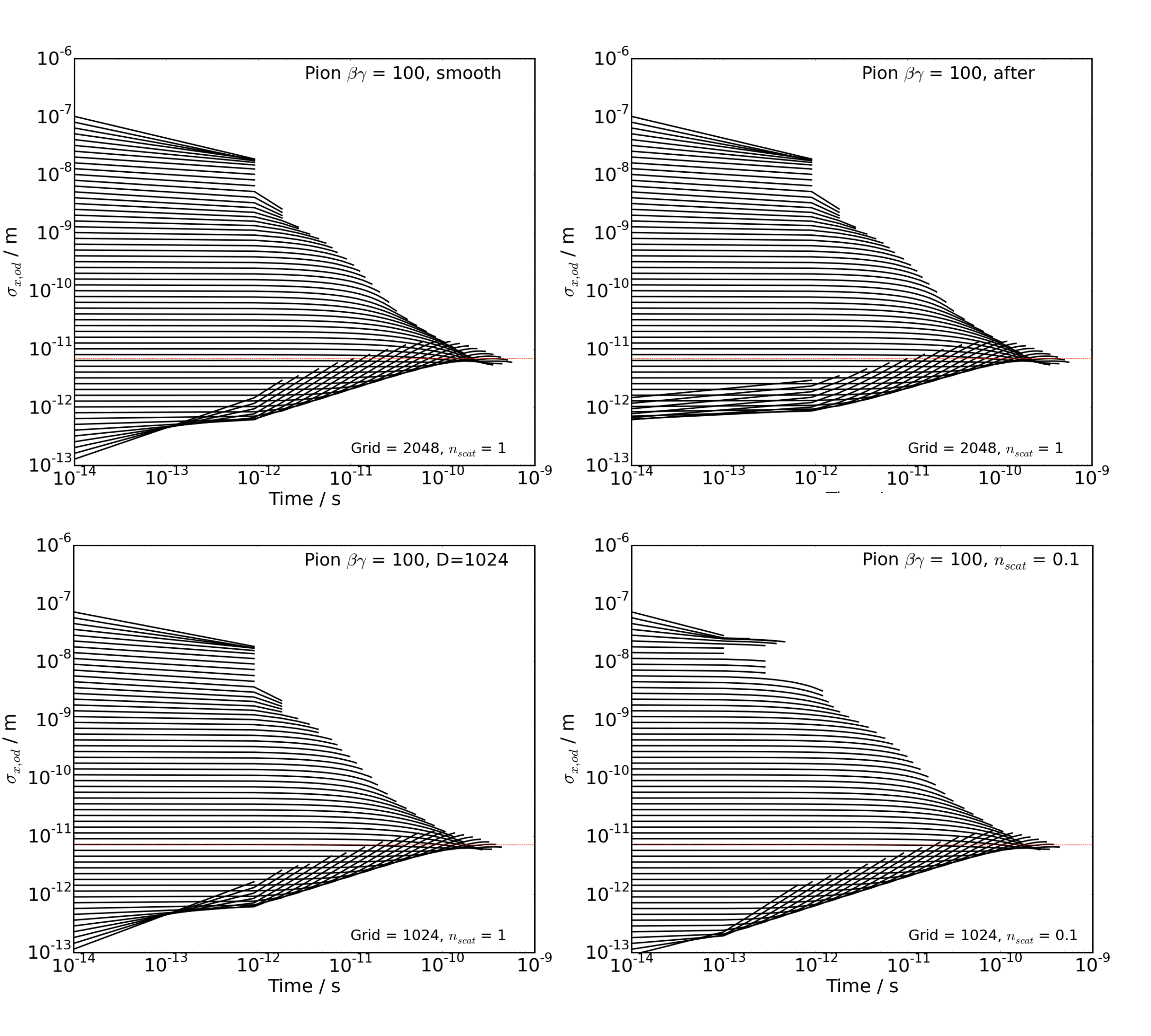}
\par\end{centering}

\caption{Cross checks of the model. Top left : baseline. Top right: measure
width after unitary evolution, rather than after scatter. Bottom left
: adjusted grid size 1024. Bottom right : $n_{scat}=0.1$ approximation.
The red line represents an identical width in each plot.\label{fig:CrossCheckPlots}}
\end{figure}

\end{widetext}

Rather than
using the Monte-Carlo technique with random
scattering times, for these tests we model a ``smoothed'' evolution, using
a constant time-per-scatter equal to the mean expected at the given
pion energy, shown in Figure \ref{fig:DecFunc}. 
This generates deterministic rather than random curves,
which demonstrate more clearly the average behavior. The baseline model
with smoothed evolution for $\beta \gamma=100$ is shown in Figure \ref{fig:CrossCheckPlots}, 
top left.  Comparison with Figure \ref{fig:EvolBG100}
  demonstrates that the smoothed evolution gives a similar approach to convergence and convergent
width to the Monte Carlo evolution, but with the advantage that it can be 
used to testing for systematics of the method independently of random
fluctuations.

Using the smoothed model for $\beta\gamma=100$
we checked that our results are not affected by the following purely
calculational adjustments. We change the point in each cycle where
the width is measured from directly after each scattering interaction
to directly after each unitary evolution. The resulting evolution
is shown in Figure \ref{fig:CrossCheckPlots}, top right. We check
that our result is not affected by the grid size, changing from a grid dimension
of 2048 to 1024, shown in Figure \ref{fig:CrossCheckPlots}, bottom
left. In the lowest energy cases for $\beta\gamma \leq 3$, to stabilize the calculation 
the Monte Carlo evolution was run using 0.1 scatters per evolution.  That is, 
unitary evolution for 0.1 $\times t_{evol}$ and application of a decoherence
function $\widetilde{P_{q}}(x_2-x_1)^{0.1}$.
We check this approximation, a weaker version of the continuum
approximation used to describe systems by master equations,
gives the same
convergent width as the default evolution. This is shown in Figure
\ref{fig:CrossCheckPlots}, bottom right. In all cases, a consistent
convergent width and a similar approach to convergence is observed.

\newpage

\end{appendix}

\bibliography{dy}

\end{document}